# File fragment recognition based on content and statistical features

Marzieh Masoumi[1] . Ahmad Keshavarz[2] . Reza Fotohi[3]

**Abstract** Nowadays, the speed up development and use of digital devices such as smartphones have put people at risk of internet crimes. The evidence of present crimes in a computer file can be easily unreachable by changing the prefix of a file or other algorithms. In more complex cases, either file divided into different parts or the parts of a file that has information about the file type are deleted, where the file fragment recognition issue is discussed. The known files are divided into different fragments, and different classification algorithms are used to solve the problems of file fragment recognition. A confusion matrix measures the accuracy of type recognition. The issue of identifying the type of file fragment due to its importance in cybercrime issues as well as antivirus has been highly emphasized and has been addressed in many articles. Increasing the accuracy in this field on the types of widely used files due to the sensitivity of the subject of recognizing the type of file under study is the main goal of researchers in this field. Failure to identify the correct type of file will lead to deviations of the results and evidence from the main issue or failure to conclude. In this paper, first, the file is divided into different fragments. Then, the file fragment features, which are obtained from Binary Frequency Distribution (BFD), are reduced by 2 feature reduction algorithms; Sequential Forward Selection algorithm (SFS) as well as Sequential Floating Forward Selection algorithm (SFFS) to delete sparse features that result in increased accuracy and speed. Finally, the reduced features are given to 3 Multiclass classifier algorithms, Multilayer Perceptron (MLP), Support Vector Machines (SVM), and K-Nearest Neighbor (KNN) for classification and comparison of the results. The proposed recognition algorithm can recognize 6 types of useful files (PDF, TXT, JPG, DOC, HTML, EXE) and may distinguish a type of file fragments with higher accuracy than the similar works done.

**Keywords** Multiclass classifier algorithms . Feature Reduction . File Fragments . File Fragment Recognition . SFS . SFFS.

✉ Marzieh Masoumi
M.maasoumi@eng.ui.ac.ir

✉ Ahmad Keshavarz
A.keshavarz@gpu.ac.ir

✉ Reza Fotohi*
R_fotohi@sbu.ac.ir; Fotohi.reza@gmail.com

[1] Faculty of Computer Engineering, University of Isfahan, Isfahan, Iran

[2] Electrical Engineering Department, Faculty of Intelligent Systems Engineering and Data Science, Persian Gulf University, Bushehr 7516913817, Iran

[3] Faculty of Computer Science and Engineering, Shahid Beheshti University, G. C. Evin, Tehran 1983969411, Iran

# 1 Introduction

Computers deal with a large number of files with different formats, which are transmitted among networks. The format of a file is an initial design of it that tells the processor devices how to organize the file information and describe their decoding algorithm in digital storage devices. The security of computers and networks reduces without the correct detection of the file type [1-4]. Detecting the file type is a significant step in adequate proceed of operating systems, firewalls, intrusion detection systems, and anti-viruses.

The need to examine files on issues such as memory analysis and Internet crimes and the dependence of these activities on data components due to the nature of disk blocks as well as data transfer in the form of network packets makes it necessary to identify the type of file components. The skill of identifying a file fragment is important in many tasks, such as trying to recover data from a corrupted drive, detecting Internet crimes, including interrupt detection, decrypting memory, reverse engineering malware, and more. Content-based analysis is needed in the absence of other identifying information to classify file segments in digital forensics. When recovering data from a damaged media, the file fragments left in the damaged media or memory may contain important information that may appear corrupted or missing in the absence of a tool to identify the type of file fragment.

The content-based algorithm includes investigating the file content and using static techniques. The contents of the file are a chain of bytes, and each byte has 256 unique characters (0-255). Therefore, the calculation of the byte pattern rate, referred to as the byte distribution rate provides a recognizable pattern for different file types.

The use of SFS and SFFS feature reduction methods for the first time to select and reduce attributes in file components is presented in this article. The proposed method uses two methods to reduce the specificity of SFS and SFFS in combination with 3 algorithms MLP, KNN, SVM as a multi-class. In this paper, a total of 6 algorithms are proposed which are a combination of feature reduction methods and classification methods, and the results of all 6 methods are fully presented along with providing the best parameters for the file fragment identification problem depending on the part size (500 and 100). An increase in classification accuracy is evident in these proposed algorithms compared to the presented algorithms.

The presented research is structured as the following. Section 2 presents related works. In Sect. 3 brings the proposed methodology. Dataset of this research is discussed in Section 4. Performance evaluation is deliberated in Section 5. Finally, the conclusion of this research is discussed in Section 6.

# 2 Related works

This section deals with a lot of research that has been done in the field of File fragment recognition in recent years.

The research done in this article and other previous articles that have provided an algorithm for identifying file parts consists of 3 general sections: reviewing feature selection techniques 4, reviewing classification techniques 5 and finally selecting the appropriate technique from each section and presenting an efficient algorithm to identify file type 3 are formed.

Feature selection techniques focus on the problem of selecting effective features from a data sample that naturally reduce the dimensions of the data in question by eliminating undesirable features. Feature selection [5] needs to be examined from two perspectives: increasing the accuracy and decreasing the number of calculations in datasets, and according to our purpose of selecting the feature as well as the specific type of our data set (e.g., data streams [6]), the appropriate approach should be taken Be.

Classification techniques, which are presented in 3 types: supervised, semi-monitored and without harm, seek to classify the members of a database into different categories according to the target factors (distance distance), and researchers are always applying classification algorithms on New problems or their combination and innovation in their structure to solve their old problems (such as the structure of simultaneous use of classification and regression) [7]

Type recognition operations include the provision of an algorithm that identifies the purpose of the problem that can identify an object in an image [7] of a disease [8] and .... using classification techniques and other new solutions such as deep learning and purpose the final is to provide an algorithm that increases accuracy and speed while reducing the execution time of the problem compared to its previous methods.

In the following, we will have an overview of the articles that have worked on the issue of file identification and have provided algorithms along with presenting the results.

McDaniel and Heidari [9] were the first to develop an algorithm for recognizing the file types based on content. Their proposed algorithms are used to generate a "fingerprint" of each file, which are detected compared with the known types, and file types. The accuracy varies between 23% and 96% depending on the algorithm used.

Li et al. [10] made slight changes to the McDaniel's model, which increased its accuracy. They provided a set of central models and used the categorization to find the minimum number of centers set with good performance while using more data patterns. This research has the accuracy of 82% (single central) and 89.5% (multi-center) with 93.5% of more sample files.

Karresand and Shahmehri [11] provided an algorithm for file fragments, which used the BFD and the standard deviation concept for file type modelling. Karresand and Shahmehri proposed the Oscar methodology for detecting the file fragments. They generated single-center printing files but used a quadratic distance metric and a norm-1 as the metric distance to compare the center with the byte frequency distribution of the file. Although Oscar recognized any file type, they reported their algorithms for JPG files using the specified pair bytes of the optimized file and the detection rate of 99.2%.

Veenman [12] extracted three features from the file's content. These features include:
- Frequency byte distribution
- The entropy obtained by frequency byte distribution of files
- The complexity of the algorithm or the Kolmogorov that uses the sequence of the substring

Fisher's linear discriminant analysis has been applied to these features to recognize the file types.

Calhoun and Coles [13] used a static algorithm and the linear FISHER one for a dataset containing 100 fragments of 2 different file types with an accuracy of 60.3% -86% (depending on the tested bytes chain). They have developed the Veenman works by the constructed classification models and presented the linear discriminant to recognize the file types. Further, they have examined machine learning algorithms to solve the data classification problem and achieved a reasonable accuracy.

Sportielo and Zanero [14] have considered a set of SVM classifications for each file type. The results of several experiments show that the features based on the byte frequency distribution have the best performance for most of the examined file types, where the SVM is very effective in distinguishing file types from the data blocks.

Gopal et al. [15] introduced the File Type Recognition (FTI) as a significant issue in digital rules and provided a systematic review of the problem, algorithmic solutions, and evaluation methodologies. They analysed the power of various algorithms in examining the files and damaged fragments. They also proposed two criteria for replacement in performance measurement as follows:
- Considering the file name extension as the correct tags (labels)
- Considering the prediction by knowledge-based algorithms in healthy files as the correct tags (labels)

The conclusion was that the SVM and KNN are better than COTS (Commercial off-the-shelf) in files where the extensions for sound files are available. Also, some COTS algorithms can detect the corrupted files by no means.

Moody and Erbacher [16] used the static analysis to recognize the file type (SADI), which includes the mean, standard deviation, average distance, standard deviation distance, and calculation of the bytes values. They used the fragments of 200 files from a dataset of 8 known files, which had a 74.2% result.

Dunham et al. [17] applied the neural networks for categorizing ten file types from a dataset, including 760 archived files with an accuracy of 91.5%.

Like et al. [18] adapted the BFD model with the Manhattan distance for comparison to determine whether the calculated files are executable or not.

Cao et al. [19] used the Gram frequency distribution and the vector space model with a 40.34% result. Ahmad et al. [16] presented two algorithms. First, they applied the cosine distance as a metric of similarity when comparing the file contents. Secondly, they divided the recognition process into two steps by Dividing and Conquering algorithms. In the first step, the similar files with the same byte frequency patterns are classified in different clusters. In the next phase, the classification, including various file types, is given to the neural networks to improve the categorization. They used 2000 different file types with the accuracy of 90.19%.

Ahmad et al. [21] also proposed two new algorithms to reduce the classification time. First, they used the Feature Selection technique and KNN classifier. The second algorithm was the sample content technique in which they used a small portion of the file to achieve the byte frequency distribution.
As described in this section, many works have been done in this approach, but then again, unfortunately, they did not specify their datasets. Moreover, they used both different types of files and datasets, which caused impossible conditions to compare them correctly with each other.

In 2015, in an experiment, Nasser Alamri [22] compared six different file types (PDF, TXT, JPG, DOC, html, and EXE) with 5 algorithms presented on the specific database, and then provided the way of comparison in future studies. We also chose Nasser Alamri's article to compare the suggested algorithms. Thus, we applied the same database and file types with Alamri [22] that provides a reasonable and fairness comparison for the present study. The purpose of this research was to recognize the file fragment types with higher accuracy than the similar research works due to the widespread use of this issue as well as its sensitivity to the correct recognizing file type. In the following, the dataset, the methodology of the proposed algorithm, and the obtained results were described. Finally, the results of this study were compared with the results of Naser Alamri study in 2015.

## 3 The proposed methodology

We provide a proposed methodology in the following section using the content and statistical features. Four phases are included in the proposed methodology: in Sect, 3.1. An overview of the proposed methodology is discussed. Sect 3.2. presents the BFD Extraction and BFD Normalization, in Sect, 3.3. the feature reduction is discussed and Sect. 3.4 presents the classification.

### 3.1 Phase 1: Overview of the proposed methodology

The train and test sets are provided by dividing a file into small fragments. Hence, we fragment complete files, but at first, we cut the header and prefix of files, which may contain information about files type. Then, we divided the rest of each file into 2 fragments of 500 and 1000 bytes, to show the effect of fragment size on the accuracy of the presented methods in the study.

As illustrated in Figure 1, SFS and SFFS algorithms were used to reduce the fragment size of the studied file and select the dynamic features. The KNN, SVM, and MLP algorithms were employed as file type detection algorithms. The LIBSVM Package was employed for SVM classification and, MATLAB Toolbox Autoencoder was utilized for the neural network.

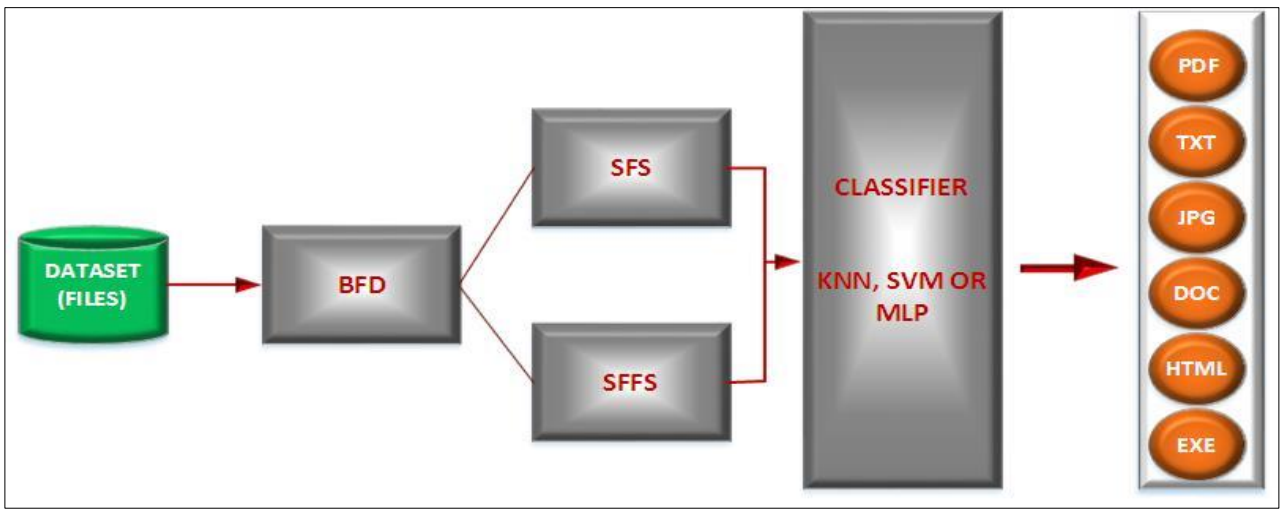

**Fig. 1** The Methodology implemented in this work

### 3.2 Phase 2: BFD Extraction and BFD Normalization

The byte frequency distribution (BFD) was used as the feature extraction algorithm. After obtaining the array bytes values rate, each member of the array was distributed by the byte frequency rate. Accordingly, the array was normalized to values between 0 and 1.

BFD is a common method for extracting attributes from a bit string, and since the received bit string is encoded, using features such as Fourier transform or wavelet transform and the like cannot be helpful. Somehow modeled with BFD can be a helpful feature in identifying the file type. On the other hand, due to the fact that high speed is required in detecting the type of file in antivirus or firewalls, the algorithm should be used as fast as possible and with appropriate accuracy. SFS and SFFS algorithms, which are among the common sequential search algorithms, have a better speed than Enumeration or random search algorithms and on the other hand, they do not have a complex parameter setting compared to algorithms such as PSO or GA. Therefore, this group of algorithms has been used in this research. In addition, the final accuracy of the test on the data set indicates the high efficiency and suitability of the proposed algorithm [23, 24].

Figure 2 displays the BFD diagram for the 500-byte fragments and Figure 3 further shows the BFD diagram for 1000-byte fragments.

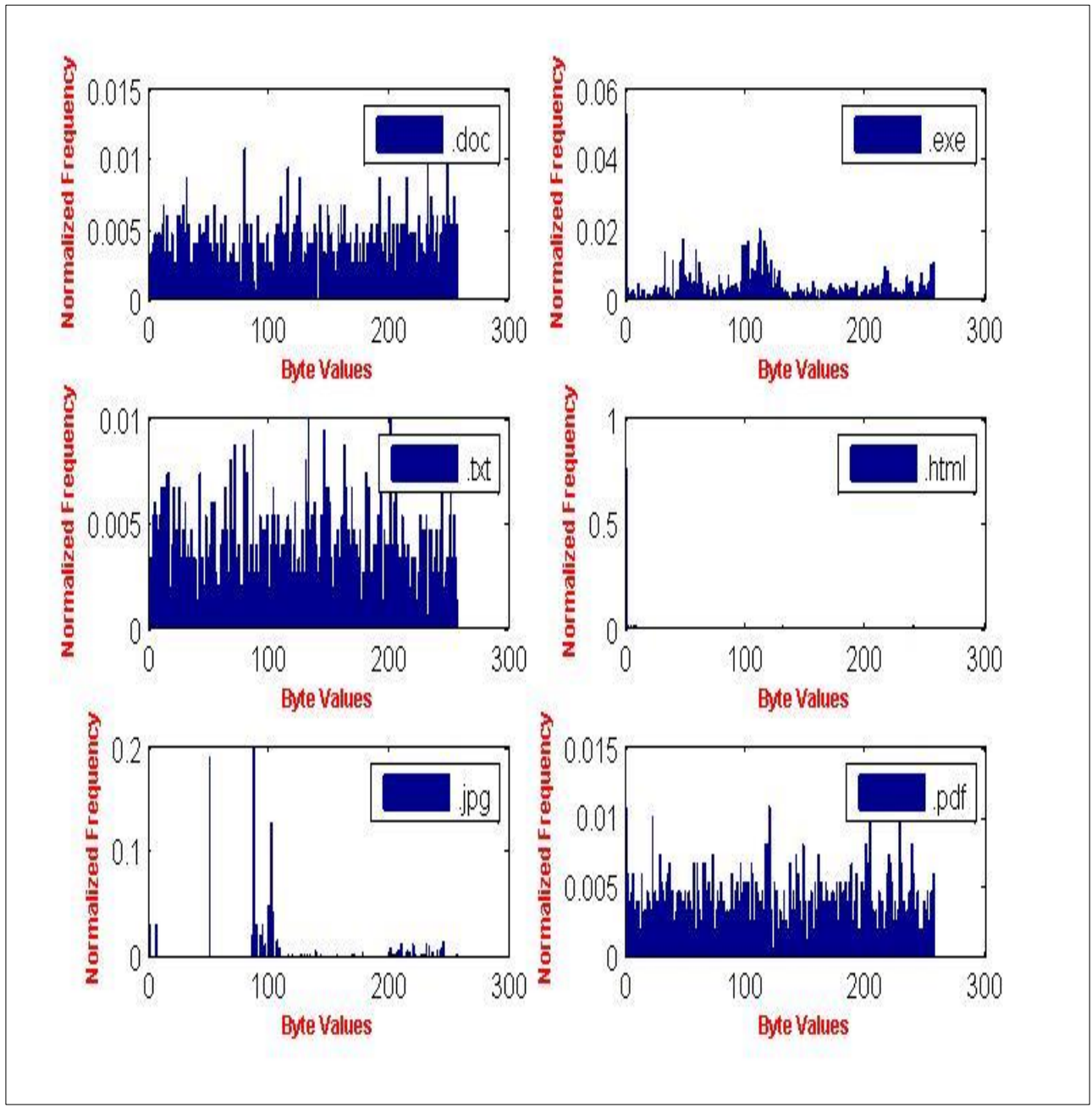

**Fig. 2** The BFD graph for the 500-fragment

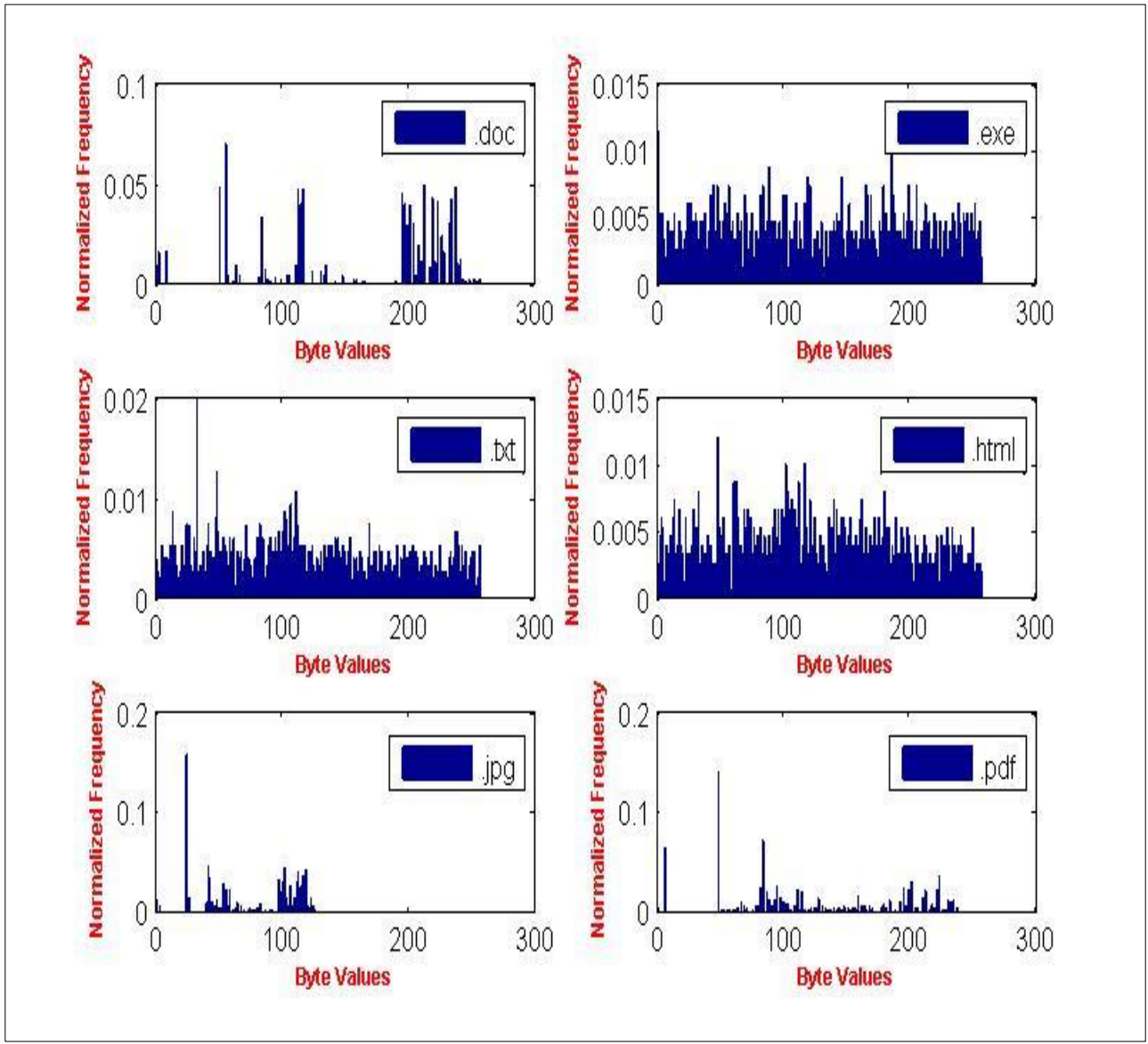

**Fig. 3** The BFD graph for the 1000-fragment

### 3.3 Phase 3: Feature Reduction

Different feature selection algorithms strain to find the best subset among the n2 candidate subsets. These algorithms increase the accuracy and speed by eliminating the outliers. We adopted SFS and SFFS algorithms in the present study as feature reduction algorithms. We tried several parameters for k, and finally, the KNN algorithm with K = 5 was considered as the criterion for feature selection. With the algorithms mentioned, 256 features, which obtain from BFD, were reduced to 24 and 39 features. We performed the feature selection process on all 500- and 1000-byte fragments. The corresponding results are given in Table 2.

**Preliminaries:** In advance of describing the corresponding algorithms formally, the following definitions have to be introduced.

Let $X_k = \{X_i : 1 \leq X_i \leq K, X_i \in Y\}$ be the set of k features from the set $Y = \{y_i : 1 \leq i \leq D\}$ of D available features.

The value $J(y_i)$ of the feature selection criterion function if only the $i_{th}$ feature $y_i = (i = 1, 2, ...)$ used will be called the individual significance $S_0(y_j)$ of the feature.

The significance $S_{k-1}(X_j)$ of the feature $X_j, j = 1, 2, ..., k$ in the set $X_k$ is defined by:

$$S_{k-1}(X_j) = J(X_k) - J(X_k - X_j) \tag{1}$$

The significance $S_{k+1}(f_j)$ of the feature $f_j$ from the set $Y - X_k$

$$Y - X_k = \{f_i : i = 1, 2, ..., D-k, f_i \in Y, f_i \neq X_t \text{ for all } x_t \in X_k\} \tag{2}$$

So, $X_k$ is defined by

$$S_{k+1}(X_j) = J(X_k + f_j) - J(X_k) \tag{3}$$

For $k = 1$ the term feature significance in the set coincides with the term of individual significance.

We shall say that the feature $x_j(b)$ from the set $X_k$ is (a) the most significant (best) feature in the set $X_k$ if:

$$S_{k-1}(X_j) = \max_{1 \leq i \leq k} S_{k-1}(X_i) \Rightarrow J(X_k - x_j) = \min_{1 \leq i \leq k} J(X_k - x_i) \tag{4}$$

(b) The least significant (worst) feature in the set $X_k$ if

$$S_{k-1}(X_j) = \min_{1 \leq i \leq k} S_{k-1}(X_i) \Rightarrow J(X_k - x_j) = \max_{1 \leq i \leq k} J(X_k - x_i) \tag{5}$$

We shall say that the feature $f_j$ from the set $Y - X_k$ is (a) the most significant (best) feature with the set $X_k$ if

$$S_{k+1}(f_j) = \max_{1 \leq i \leq D-k} S_{k+1}(f_i) \Rightarrow J(X_k + f_j) = \max_{1 \leq i \leq D-k} J(X_k - f_i) \tag{6}$$

(b) The least significant (worst) feature concerning the set $X_k$ if

$$S_{k+1}(f_j) = \min_{1 \leq i \leq D-k} S_{k+1}(f_i) \Rightarrow J(X_k + f_j) = \min_{1 \leq i \leq D-k} J(X_k - f_i) \tag{7}$$

**Sequential Forward Selection (SFS) Algorithm:** In the "sequential feature selection" (SFS) algorithm, the process starts with an empty set. Then, in each repetition, a feature is added to the answer set by employing the evaluation function used. This is repeated until the selection of the required features [25]. Using SFS, we achieved 24 features and 36 for 500-byte and 1000-byte fragments, respectively. The Sequential Forward Selection (SFS) Algorithm SFS is shown in Figure 4.

---

**Algorithm SFS:** Pseudo code for Sequential Forward Selection (SFS) Algorithm

1: ***Procedure*** Sequential Forward Selection (SFS)

2:    **Start** with an empty set $y_0 = \{0\}$

3:    **Choose** the next best features

4:    $x^+ = \arg_{x \notin y_k} \max_j (y_k + x)$

5:    **Update** set

6:    $x^+ = \arg_{x \notin y_k} \max_j (y_k + x) s$

7:    **Return** to Step 3

4: ***End Procedure***

---

**Fig. 4** The Sequential Forward Selection (SFS) Algorithm

**Sequential Floating Forward Selection (SFFS) Algorithm:** First, the sequential floating forward selection (SFFS) algorithm begins with an empty set of features. For each step, the best feature that satisfies the criterion function is placed in the current set. That is, one stage of the sequential forward selection is performed. The SFFS progresses with dynamic increasing or decreasing of the feature numbers to achieve the optimal number of them [26]. Using SFFS, we obtained 36 and 39 features for the 500-byte and 1000-byte fragments, respectively. The Sequential Floating Forward Selection (SFFS) Algorithm SFFS is shown in Figure 5.

---

**Algorithm SFFS:** Pseudo code for Sequential Floating Forward Selection (SFFS) Algorithm

1: ***Procedure*** Sequential Floating Forward Selection (SFFS)

2:    **Start** with an empty set $y_0 = \{0\}$

3:    **Choose** the next best features

4:    **Update** set

5:    $k = k+1, y_{k+1} = y_{k+1} + x^+$

5:    **Choose** the worst features

6:    $x^- = \arg_{x \notin y_k} \max_j (y_k - x)$

7:    If $j(y_k - x) \succ y_k$

8:    $k = k+1, y_{k+1} = y_k - x^-$

9:    **Return** to Step 3

7:    Else

8:    **Return** to Step 2

4: ***End Procedure***

---

**Fig. 5** Sequential Floating Forward Selection (SFFS) Algorithm

## 3.4 Phase 4: Classification

At the stage of categorizing the type of file fragments, the acquired features are used as inputs in three algorithms, KNN, SVM, and MLP as described below.

The KNN algorithm is a simple supervised algorithm that stores all available cases in different categories based on a similarity measure and classifies new cases [27]. The k parameter displays the number of closest neighbors in the feature space. We used a KNN algorithm with k = 4, 6, 8, and 10; the results are illustrated in table 3 for 1000-byte and table 5 for 500-byte fragments.

The SVM algorithm is a supervised algorithm that performs classification by finding the hyperplane, which maximizes the margin between the two classes [27]. In this study of file fragment recognition, we use SVM algorithm as the second classification approaches with Radial Basis Function (RBF) kernel as well as a different c parameter, c = 0.1, 0.2, 0.3, as shown in table 3 for 1000-byte and table 5 for 500-byte fragments.

The MLP is the third classification algorithm used in the study. It is a type of feedforward neural network, which may differentiate data that is not linearly separable [27]. We use MLP with 1 hidden layer and sigmoid activation function as shown in Figure 6 and the result in Table 3 for 1000-byte and table 5 for 500-byte fragments.

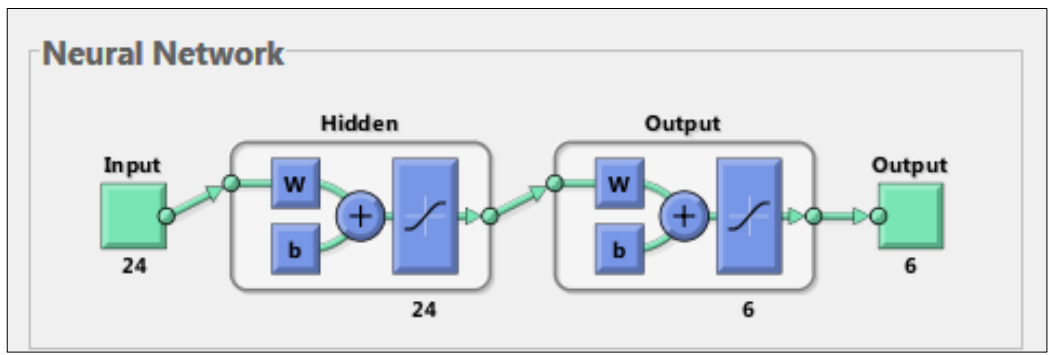

**Fig. 6** The MLP used model in this study

We tried several parameters in each algorithm to obtain the best result. The corresponding results are given in Table 4 for 1000-byte and Table 6 for 500-byte fragments.

The classifiers presented in this article classify the results into 6 categories (PDF, TXT, JPG, DOC, HTML, EXE) which are widely used file types as 6 class classifiers and the results presented in this article are related to the multi-class mode. The results presented in this article are the result of 10 repetitions of experiments with different parameters and the ratio of training data to test data in the form of 70.30, 80.20 and 90.90 to evaluate the effect of changing the percentage of test and the most, which, as expected, the best results in each section it is related to the ratio of 90/90. (Table 5) In the MLP method, the lowest accuracy for 1000-byte parts is related to the training rate of 70.30 with a result of 92% for MLP, 91% for KNN (k = 10) and 94% for SVM (c = 0.1). (Table 6) In the MLP method, the lowest accuracy for 500-byte parts is related to the training rate of 70.30 with a result of 92% for MLP, 94% for KNN (k = 10) and 94% for SVM (c = 0.1).

# 4 Dataset

Most published research in this area has used data that is often not clearly described and is not publicly available, and as a result, it is not possible to directly apply different approaches to each other or to new techniques that address this issue. Focused have been compared. Some researchers compared their approaches to different datasets, but the results of one research project were not consistent with another. As a result, one journal may present evidence that one particular approach performs better than the other (on their data), and another journal may present conflicting results with the previous journal. Our goal is to select the same data set with the background paper to provide a fair framework for comparison with other approaches to file type identification to assess the accuracy of their classification with the same types of file types that in Nasser Al-Amri article in exactly the same conditions run on this data set and their results are given for comparison.

The standardized Govdocs1 dataset, containing 1,000 lists of 1,000 content files, was used in this research. From 3 random folders of this database, we extracted 100 files from each sample of TXT, JPG, HTML, and PDF (totally 600 files) with a minimum size of 4Kb. The EXE files were obtained from Windows system files by considering their minimum size. The data applied in the program are standard data that are used extensively in similar studies; these are available at the following address: http://digitalcorpora.org/corpora/govdocs.

In the present research, we focused on the file types included in the dataset section; the statistical descriptions are given in Table 1.

**Table 1:** Files types

| Type | Number | Average size | Maximum size | Minimum size |
|------|--------|--------------|--------------|--------------|
| DOC  | 100    | 345,796      | 9,023,488    | **12,800**   |
| EXE  | 100    | 187,498      | 6,440,448    | **4,724**    |
| TXT  | 100    | 391,526      | 1,063,025    | **4,061**    |
| HTML | 100    | 76,370       | 16,497,395   | **4,008**    |
| JPG  | 100    | 162,012      | 7,778,639    | **4,023**    |
| PDF  | 100    | 608,778      | 10,891,418   | **4,710**    |

# 5 Performance evaluation

The proposed method performance is assessed in the following section.

### 5.1 Simulation results

In this section, the results obtained from the implementation are analyzed, and finally, the result of the proposed algorithm is compared with other available algorithms. We presented the results of the implementation of the proposed solution in two parts of 1000-byte and 500-byte fragments. For both 1000-byte and 500-byte fragments, we reduced the features obtained from BFD components via SFS and SFFS algorithms. At that point, we gave the reduced set of features to the SVM, KNN, and MLP Multi class classifier algorithms. The accuracies of the multi class classifiers are given in the tables. The results presented in the tables are the outcomes of 10 repetitions of the algorithm with various parameters. The best result of each classifier algorithm with a different combination of the training rate and the corresponding parameters of the feature reduction algorithm for 1000-item fragments are shown in Table 3.

**Table 2:** Feature reduced results

| Algorithm | The Number of features selected For 1000 fragments | The Number of features selected For 500 fragments |
|---|---|---|
| SFS | 36 | **24** |
| SFFS | 39 | **35** |

**Table 3:** Results of 1000 fragments ()

| Parameter | Multi class Classifier | Algorithm | Number of features | Train/Test | Accuracy |
|---|---|---|---|---|---|
|  | MLP | SFFS | 39 | 90/10 | 95% |
| K=4 | KNN | SFS | 36 | 90/10 | 96% |
| K=6 | KNN | SFS | 36 | 90/10 | 97% |
| K=8 | KNN | SFFS | 39 | 90/10 | 97% |
| K=10 | KNN | SFS | 36 | 90/10 | 97% |
| C=0.1 | SVM | SFFS | 39 | 90/10 | 97% |
| C=0.2 | SVM | SFFS | 39 | 90/10 | 98% |
| C=0.3 | SVM | SFFS | 39 | 90/10 | 98% |

The best results obtained in 1000-byte fragments with the best possible combinations are given in Table 4 below.

**Table 4:** The best results of 1000 fragments

| Multi class Classifier | algorithm | Number of features | Train/test | accuracy |
|---|---|---|---|---|
| MLP | SFFS | 39 | 90/10 | 95% |
| KNN | SFS and SFFS | 36 or 39 | 90/10 | 97% |
| SVM | SFFS | 39 | 90/10 | 98% |

According to Table 4, the MLP algorithm with 96% accuracy, the KNN algorithm with an accuracy of 97%, and the SVM algorithm with an accuracy of 98% completed their process in the 1000-byte fragments. Accordingly, the SVM algorithm is considered the best algorithm for recognizing the 1000-byte files with an accuracy of 98%.

The best result of each classifier algorithm with a different combination of the training rate and the corresponding parameters of the feature reduction algorithm for 500-item fragments are shown in Table 5.

**Table 5:** Results of 500 fragments

| Parameter | Multi class Classifier | Algorithm | Number of features | Train / Test | Accuracy |
|---|---|---|---|---|---|
|  | MLP | SFFS | 35 | 90/10 | 96% |
| K=4 | KNN | SFFS | 35 | 90/10 | 98% |
| K=6 | KNN | SFFS | 35 | 90/10 | 98% |
| K=8 | KNN | SFFS | 35 | 90/10 | 98% |
| K=10 | KNN | SFFS | 35 | 90/10 | 98% |
| C=0.1 | SVM | SFFS | 35 | 90/10 | 98% |
| C=0,2 | SVM | SFFS | 35 | 90/10 | 97% |
| C=0,3 | SVM | SFFS | 35 | 90/10 | 98% |
| C=0,4 | SVM | SFFS | 35 | 90/10 | 98% |

In the MLP method, the lowest accuracy for 1000-byte parts is related to the training rate of 70.30 with a result of 92% for MLP, 91% for KNN (k = 10) and 94% for SVM (c = 0.1).

The best results obtained in 500-byte fragments with the best possible combinations are given in Table 6.

**Table 6:** The best results of 500 fragments

| Multi class Classifier | Algorithm | Number of features | Train / Test | Accuracy |
|---|---|---|---|---|
| MLP | SFFS | 35 | 90/10 | 96% |
| KNN | SFS & SFFS | 36 or 39 | 90/10 | 98% |
| SVM | SFFS | 39 | 90/10 | 98% |

According to Table 6, the MLP algorithm with 95% accuracy, the KNN algorithm with an accuracy of 98%, and the SVM algorithm with an accuracy of 98% completed their process in the 500-byte fragments. Accordingly, the KNN and SVM algorithms are considered the best algorithms for recognizing the 500-byte files with an accuracy of 98%.

According to Table 6 in the MLP method, the lowest accuracy for 500-byte components is related to the training rate of 70.30 with a result of 92% for MLP, 94% for KNN (k = 10) and 94% for SVM (c = 0.1).

## 5.2 Analysis of the research results

The best results of the research by comparing two SFS and SFFS algorithms, as well as both 500-byte and 1000-byte fragments, are presented in Table 7.

**Table 7:** Results to be compared

| Multi class Classifier | Number of features | Fragment size | Train / Test | Accuracy |
|---|---|---|---|---|
| MLP-s | 35 | 500 Byte | 90/10 | 96% |
| K-NN-s | 35 | 500 Byte | 90/10 | 98% |
| SVM-s | 35 | 500 Byte | 90/10 | 98% |

Referring to Table 7, the MLP algorithm provides its best result on the 500-byte fragments with SFFS feature reduction algorithm by selecting 35 features. The best result recorded for the MLP algorithm in this study is 96%. The KNN algorithm also provides its best result on the 500-byte fragments with the SFFS feature reduction algorithm by selecting 35 features. The best result recorded for the KNN algorithm in the current study is 98%. The SVM algorithm also provides its best result on the 500-byte fragments with the SFFS feature reduction algorithm by selecting 35 features. The best result recorded for the SVM algorithm in this research is 98%. We called these proposed algorithms SVM-s, KNN-s, and MLP-s, respectively.

As specified by the results, by increased length of the fragments from 500 to 1000 bytes, the examined algorithms provide either weaker or similar results with a minimal alteration, which can be due to a small difference in the number of features obtained from SFS and SFFS reductions algorithms for 1000-bytes fragments compared to 500-byte fragments.

As illustrated in Table 7, the SVM and KNN algorithms with similar accuracy of 98% are at the highest place, and the MLP algorithm with an accuracy of 96% occurs in a lower place. This means feature reduction by SFFS algorithm will provide better results than the SFS algorithm for 1000-byte and 500-byte fragments. Moreover, the SVM and KNN algorithms have a better performance than the MLP algorithm.

## 5.3 Comparison of the proposed algorithm with other algorithms

The study in the field of recognizing the file type includes a large number of file types as well as different databases. This leads to complexity in the comparison and conclusion of the research. In 2015, in an experiment, Nasser Alamri selected 6 different file types (PDF, TXT, JPG, DOC, html, and EXE) and reduced the features via the PCA feature reduction. Then and there, he compared the reduced features set with 5 algorithms of SVM, KNN, the neural network based on the core function radius, the neural network with perceptron core, and linear discriminant analysis on the same database. The relevant database has randomly extracted the sample data from the Govdoc dataset, and 100 samples were taken from each file of which the subsets are also randomly extracted. The results are shown in Table 8. We also matched a variety of file types with the files provided to compare our work with another research.

**Table 8:** Results obtained in Alamri′s 2015 paper

| Multi class Classifier | Number of features | Fragment | Train/test | Accuracy |
|---|---|---|---|---|
| LDA | 64 | 500 Byte | 90/10 | 93% |
| SVM | 64 | 500 Byte | 80/20 | 94% |
| K-NN | 8 | 500 Byte | 90/10 | 97% |
| NN-RBF | 4 | 1000 Byte | 80/20 | 88% |
| NN-MLP | 64 | 500 Byte | 90/10 | 94% |

As shown in Table 8, the KNN algorithms with the accuracy of 97% and the NN-RBF algorithm with an accuracy of 88% have the least accuracy in the Nasser Alamri paper. Figure 7 shows the comparison between our proposed research algorithms and the Alamri's paper.

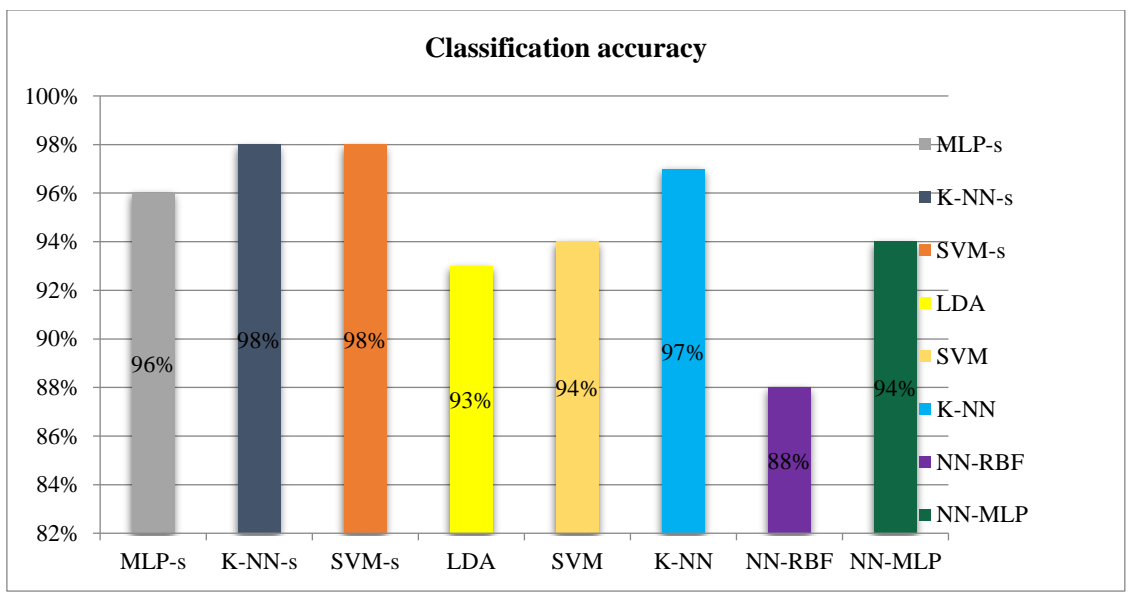

**Fig. 7** Comparison of the proposed algorithm of this research with the results of Alamri's paper (2015)

In Figure 7, the MLP-s column describes the MLP classification algorithm by SFFS feature reduction approach, and the K-NN-s column represents the K-NN classification algorithm by the same approach. Further, the SVM-s column represents the SVM classification algorithm by SFFS feature reduction approach. According to Figure 7, the KNN-s, SVM, and MLP-s approaches by respectively 1%, 4%, and 2% increase in the accuracy rate show the increasing trend in the accuracy of this research compared to Alamri's. Also, KNN and SVM algorithms combined with the SFFS feature reduction approach indicate the highest accuracy of the categorization (98%) among the eight algorithms examined.

## 6  Conclusion

File type's detection is an essential task for many security programs. Although there are lots of programs to deal with detection of computer file types, there are just minimal algorithms for detecting them. However, the primary issue in detecting the file type is the classification of the file fragments since there are no headers (a part of the file containing information about the file type) or systemic file information, which can specify the file type. The general algorithm to classify file fragments is to examine the histogram of its byte frequency and sometimes analyze other statistics obtained. The statistical distance between the histogram and the known distributions of different files types can be calculated, which will be used to distinguish different data types. Based on recent research, although the classification of file fragments in many common file types can be done with high accuracy, this algorithm has some limitations to detect the type of file, running time and accuracy. A higher degree of accuracy obtained in this study compared with previous studies. In this paper, the problem of recognizing the file fragment was begun by considering 1000-byte and 500-byte fragments of each file. The BFD algorithm extracted the features of each file fragment. Then, by two SFS and SFFS feature reduction algorithms, the features extracted from each fragment were reduced to 24-39 features depending on the length of the file fragment. The reduced features were considered as inputs of three MLP, KNN, and SVM classification algorithms to obtain the accuracy of the classification algorithms. The best result in this study was achieved as 98%.

## Conflict of Interest

The authors declare that they have no conflict of interest.

## DATA Availability Statement

The data of this paper is the result of simulation and all the data are presented in the form of graphs inside the paper. There is no private data in this article.

## Funding

None

## Informed consent statement

None